\documentclass[
  reprint,
  amsmath, amssymb,
  aps, prb,
  superscriptaddress,
  nofootinbib,
  floatfix
]{revtex4-2}

\usepackage{graphicx}
\usepackage{bm}
\usepackage{booktabs}
\usepackage[colorlinks=true, allcolors=black]{hyperref}
\usepackage[utf8]{inputenc}
\graphicspath{{figs/}}

\newcommand{\VB}{V_{\mathrm{B}}^{-}}
\newcommand{\hBN}{hBN}
\newcommand{\us}{\,\mu\mathrm{s}}
\newcommand{\ns}{\,\mathrm{ns}}

\begin{document}

\title{Tuning Boron-Vacancy Qubit Coherence through Layer Number in hBN}

\author{Juan M. Florez}
\email{jflorezu.docente@ucentral.cl, jmflorez12@gmail.com}
\affiliation{Facultad de Ingeniería y Arquitectura, Universidad Central de Chile, Santiago, Chile}

\author{Eric Su\'arez Morell}
\email{eric.suarez@usm.cl}
\affiliation{Grupo de Simulaciones, Departamento de F\'isica, Universidad
T\'ecnica Federico Santa Mar\'ia, Valpara\'iso, Chile}

\date{\today}

\begin{abstract}
The negatively charged boron vacancy ($\VB$) in hexagonal boron nitride
(\hBN{}) was the first optically addressable spin qubit identified inside a van
der Waals crystal, allowing atomically defined placement relative to a target.
Its coherence in bulk \hBN{} is limited by the boron nuclei of the layers
flanking the defect plane, which a thin flake removes. Here, we use a
generalized cluster-correlation expansion with an extended central-spin block to
calculate the Hahn-echo coherence time of $\VB$ in h$^{11}$B$^{15}$N as a
function of layer number. Our results show that $T_2$ rises from $199\ns$ in the
bulk limit to $653\ns$ in a monolayer, a factor of $3.3$ that is already
saturated at three layers and that a sublattice decomposition attributes
entirely to boron. The enhancement is confined to low field, is insensitive to
stacking registry and twist angle, and requires alignment to within about two
degrees, an onset we reproduce with no free parameters. Under dynamical
decoupling a $1/e$ threshold returns $206\ns$ for every thickness, but this is a
zero of the first-shell modulation rather than a decay: measured without a
threshold, the layer contrast survives and reaches a factor of $170$. Layer
number thus emerges as a design parameter for $\VB$-based sensing.
\end{abstract}

\maketitle

\section{Introduction}
\label{sec:intro}

The negatively charged boron vacancy $\VB$ in hexagonal boron nitride (\hBN{})
was the first optically addressable spin defect identified in a van der Waals
crystal~\cite{Gottscholl2020}, and it now supports room-temperature optically
detected magnetic resonance and coherent control~\cite{Gottscholl2021SciAdv}, sensing of temperature, pressure, strain and magnetic
field~\cite{Gottscholl2021NC}. Other optically addressable spin defects have
since been identified in \hBN{} and in other van der Waals
hosts~\cite{Stern2022,Stern2024,Scholten2024}. Its advantage, for instance, over the nitrogen--vacancy centre in
diamond is structural rather than spectroscopic: it resides inside a layered
crystal that can be exfoliated and stacked, so the distance between sensor and
target is set by the number of atomic layers rather than by implantation depth.
The electronic structure of the defect (an $S=1$ triplet with zero-field
splitting $D = 3.47$~GHz and spin-selective optical transitions) has been
established through ab initio theory~\cite{Ivady2020}. That same structure is often described as its liability. Both sublattices of
\hBN{} are essentially fully spinful ($^{14}$N, $I=1$; $^{11}$B, $I=3/2$;
$^{10}$B, $I=3$), so no isotopic dilution can empty the nuclear bath, and
measured Hahn-echo coherence times at low field are of order
$100\ns$~\cite{Haykal2022,Rizzato2023,Ramsay2023}.\\

The mechanism behind that value has recently been worked out in detail for
\emph{bulk} \hBN{} by T\'ark\'anyi and Iv\'ady~\cite{Tarkanyi2026}, who mapped
the coherence from $0$ to $3$~T and identified five distinct field regions.
Their conclusion for the low-field regime is the starting point of this paper:
the coherence-limiting noise is dominated by ``boron nuclear spins located in
the layers above and below the $\VB$ center's plane,'' while the strongly
coupled in-plane nitrogens contribute only an envelope modulation. The reason
the in-plane nuclei are quiet is a symmetry selection rule, identified first
for near-plane nuclei of the nitrogen--vacancy centre in
diamond~\cite{Nizovtsev2018} and highlighted for planar defects in
two-dimensional hosts~\cite{Seo2024}: for a planar defect with $B_0 \parallel c$, the
horizontal mirror forces the transverse hyperfine components $A^{zx}$ and
$A^{zy}$ to vanish identically at every nucleus lying \emph{in} the defect
plane. General reviews of defects in \hBN{} for quantum technologies have
recently surveyed the broader landscape of such
centres~\cite{Vogl2026}.\\

It is then natural to ask what happens when the layers above and below are
simply absent. A monolayer places every nucleus of the bath in the mirror plane, so the
dominant low-field noise source identified in the bulk is removed by
construction. Earlier cluster expansions anticipated a thickness dependence at
high field: Ye, Seo and Galli found a factor of two between monolayer and bulk
for a model defect in natural \hBN{} at $0.5$~T, and predicted that isotopic
purification would sharpen the contrast~\cite{Ye2019}; Lee, Park and Seo
obtained a factor of $1.7$ for $\VB$ itself at $3$~T~\cite{Lee2022},
attributing it to the smaller number of bath spins. Both calculations were
deliberately performed in the strong-field limit, where the electron can be
frozen along $z$; the generalized expansions required at low field
(Sec.~\ref{sec:method}) have so far been applied to bulk crystals
only~\cite{Tarkanyi2026,Lee2026}. Here, we compute the Hahn-echo coherence of
$\VB$ as a function of the number of \hBN{} layers in the low-field regime
relevant to current experiments, in isotopically purified h$^{11}$B$^{15}$N,
and find a factor of $3.3$ between a monolayer and the bulk limit. We then
show, by separating the two sublattices, that the effect
is carried entirely by boron, in agreement with the bulk analysis; that
it closes between $25$ and $85$~mT and reopens above the ground-state level
anticrossing, holding a factor of $2$--$3$ from $220$ to $600$~mT; that it is
destroyed by a few degrees of field misalignment; and that it survives
dynamical decoupling, although a threshold-based coherence time conceals that
fact.\\

The paper is organized as follows. Section~\ref{sec:model} defines the model, the qubit, the coherence time, and the layer geometry. Section~\ref{sec:method} details the generalized cluster-correlation method and the extended central spin block. Section~\ref{sec:results} presents the results for the layer-number enhancement, its field dependence, alignment tolerance, and stacking sensitivity. Section~\ref{sec:limits} discusses the limitations of our calculation. Section~\ref{sec:conclusions} summarizes our conclusions and outlines the experimental requirements for testing the prediction.

\begin{figure*}[t]
\includegraphics[width=\textwidth]{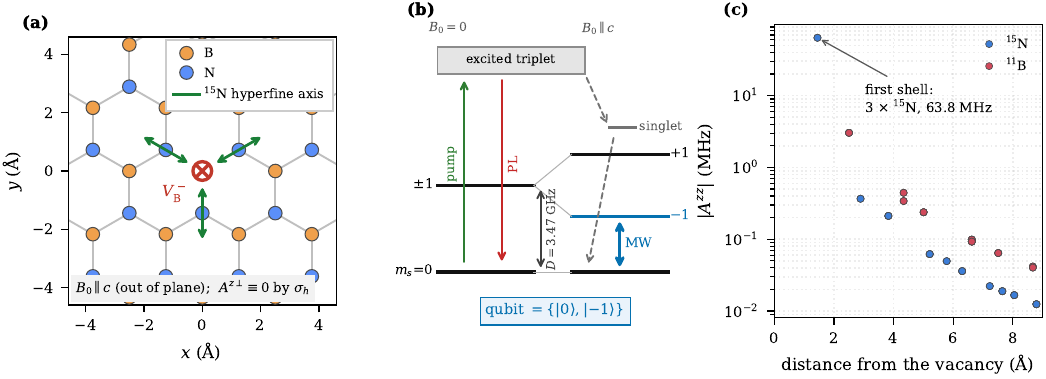}
\caption{The defect. (a) A boron vacancy in a \hBN{} layer; every lattice site
carries a nuclear spin, so the bath cannot be emptied by isotopic dilution.
Green arrows mark the principal hyperfine axis of the three first-shell
nitrogens, which lies \emph{in the plane} along the dangling bond rather than
along $c$~\cite{Gracheva2023}, so that the component entering $A^{zz}$ is
$A_\perp$ and not $A_\parallel$ (Sec.~\ref{sec:hamiltonian}). (b) The qubit
levels: the $S=1$ ground state split by $D$ at zero field and Zeeman-split by
$B_0$, the $\lvert 0\rangle \leftrightarrow \lvert -1\rangle$ pair being the
qubit. Schematic, not to scale. (c) Secular hyperfine couplings by shell for the
simulated bath, h$^{11}$B$^{15}$N. $A^{zz}$, the shift in a nucleus's
precession frequency when the qubit changes state, is defined in
Sec.~\ref{sec:hamiltonian}; the three first-shell nitrogens, at $63.8$~MHz,
dominate by more than an order of magnitude. Every in-plane nucleus has
$A^{z\perp} = 0$ exactly, by the mirror symmetry drawn in panel (a).}
\label{fig:system}
\end{figure*}

\section{Material Model}
\label{sec:model}

\subsection{The qubit}
\label{sec:qubit}

Removing a boron atom from the \hBN{} lattice and adding an electron leaves
three nitrogen dangling bonds sharing two unpaired electrons with parallel
spins, so the ground state is an $S=1$ triplet. The zero-field splitting
$D = 3.47$~GHz separates the $m_s = 0$ sublevel from the $m_s = \pm 1$ doublet,
and an applied field along $c$ splits the doublet by the electron Zeeman energy.
Optical pumping through a spin-selective intersystem crossing prepares the
defect in $m_s = 0$ and makes the photoluminescence spin-dependent, which is
what allows initialisation and readout with light. Microwaves resonant with the
$m_s = 0 \leftrightarrow m_s = -1$ transition drive that pair coherently, as
described for the VB$^-$~\cite{Ivady2020}. The qubit is therefore not the whole triplet: it is the two-level system
$\{\lvert 0\rangle, \lvert -1\rangle\}$ embedded in it, with $\lvert +1\rangle$
left far off resonance by $D$. Figure~\ref{fig:system} shows the defect, these
levels, and the hierarchy of nuclear couplings the lattice presents to them.

\subsection{Coherence time}
\label{sec:t2}

A qubit is useful only while it keeps the relative phase between its two states.
That phase is scrambled by the fluctuating magnetic field of the nuclear bath.
The Hahn echo removes the part of the scrambling that is static: a $\pi$ pulse
applied halfway through the experiment reverses the accumulated phase, so any
field that did not change during the sequence cancels exactly. What survives is
the effect of the bath \emph{evolving} during the sequence, and the time over
which the echo amplitude falls to $1/e$ of its initial value is the Hahn-echo
coherence time $T_2$. Everything a spin qubit does is bounded by it. The number of gates is roughly
$T_2$ divided by the gate time, and for sensing the shot-noise-limited
sensitivity improves as $\sqrt{T_2}$. 
Two cautions apply throughout. First, $T_2$ is a threshold crossing, and for
$\VB$ the echo is deeply modulated rather than smoothly decaying, so the crossing
must be located on a time grid fine enough to resolve the modulation
(Sec.~\ref{sec:conv}). Second, our calculation includes only nuclear-spin-induced
decoherence; other paramagnetic defects, spin--lattice relaxation and pulse
imperfections can only shorten $T_2$, so our numbers are upper bounds, as are
those of Ref.~\cite{Tarkanyi2026}.

\subsection{About layer number}
\label{sec:layers}

There is exactly \emph{one} vacancy in the system, and
it sits in the plane we label layer $0$. All other layers are pristine \hBN{};
they contain no defect and contribute nothing to the qubit itself. The qubit
exists, with the same electronic structure, whether or not those layers are
present. What the additional layers contribute is nuclear spins. We stack them in the
AA$'$ registry of real bulk \hBN{}, boron over nitrogen and nitrogen over
boron, the basis swapped on every odd layer, at the experimental interlayer
separation $3.33$~\AA{}, with \emph{zero twist angle and zero interlayer slide}.
The stack grows outward from the defect plane one side at a time: $\{0\}$ for a
monolayer, $\{0,+1\}$ for a bilayer, $\{-1,0,+1\}$ for a trilayer,
$\{-2,\dots,+2\}$ for five. Twisted stacks and the slid registries responsible for
interfacial ferroelectricity move the out-of-plane nuclei; we take AA$'$ as the
reference and show in Sec.~\ref{sec:res-stack} that the results are insensitive
to both, at the $0.2\%$ level.\\

We have verified the following facts:
(1) The first shell is untouched. The three nearest nitrogens sit at
$1.446$~\AA{} with the same hyperfine tensor in every configuration, monolayer
through bulk: $A^{zz} = 64$~MHz for the $^{15}$N of h$^{11}$B$^{15}$N, the
value measured for $^{14}$N, $45.5$~MHz~\cite{Gracheva2023}, scaled by the
nuclear gyromagnetic ratio $1.403$ and consistent with its direct measurement
on $^{15}$N-enriched crystals~\cite{Sasaki2023}. Nothing about the strongly
coupled core changes with layer number.
(2) In-plane nuclei are protected in every configuration. For all
nuclei in layer $0$ we find $A^{zx} = A^{zy} = 0$, in the
monolayer and in every slab, as the mirror selection rule
requires~\cite{Seo2024}.
(3) Out-of-plane nuclei are not protected, and adding a layer below
does not fix it. The mirror maps an in-plane nucleus onto \emph{itself}, which
is why its transverse coupling must vanish; it maps an out-of-plane nucleus onto
a \emph{partner} in the opposite layer, which imposes no such constraint on
either one. In a symmetric trilayer we find $\max|A^{z\perp}| = 0.378$~MHz in
layer $+1$ and the same in layer $-1$.
The monolayer is thus the only geometry in which every nucleus of the bath is
protected.\\

With two layers $\{0,+1\}$, the slab has no horizontal mirror, so the second item above does not apply. It nonetheless holds down to the working precision of our calculation, because the spin density is confined to the defect plane and the dipolar field of a source
at $z = 0$ evaluated at a nucleus at $z = 0$ has no out-of-plane component
whatever sits above. A first-principles treatment, in which the defect orbital
acquires a little weight on the adjacent layer, would give the bilayer a small
non-zero $A^{z\perp}$ in plane \cite{Tarkanyi2026,Seo2024}. We expect the correction to be minor, since the bilayer is already within $16\%$ of the trilayer; the vanishing is a property of our model rather than a protection enforced by the crystal, and only the monolayer and the symmetric slabs obey the exact rule.

\subsection{Spin Hamiltonian}
\label{sec:hamiltonian}

The electron spin $\mathbf{S}$ and the nuclear spins $\mathbf{I}_i$ obey
\begin{equation}
\begin{aligned}
H = {}& D\!\left(S_z^2 - \tfrac{2}{3}\right) + \gamma_e \mathbf{B}_0\!\cdot\!\mathbf{S}
     + \sum_i \mathbf{S}\!\cdot\!\mathsf{A}_i\!\cdot\!\mathbf{I}_i \\
     &+ \sum_i \left[ -\gamma_i \mathbf{B}_0\!\cdot\!\mathbf{I}_i
        + \mathbf{I}_i\!\cdot\!\mathsf{P}_i\!\cdot\!\mathbf{I}_i \right]
     + \sum_{i<j} \mathbf{I}_i\!\cdot\!\mathsf{J}_{ij}\!\cdot\!\mathbf{I}_j .
\end{aligned}
\label{eq:H}
\end{equation}
$\mathsf{A}_i$ is the hyperfine tensor, $\mathsf{P}_i$ the nuclear quadrupole
tensor and $\mathsf{J}_{ij}$ the internuclear dipolar coupling. The component
$A^{zz}$ is the shift in nucleus $i$'s precession frequency when the qubit
changes state, and is the quantity plotted by shell in
Fig.~\ref{fig:system}(c); the transverse components $A^{zx}, A^{zy}$ tilt the
nuclear quantisation axis differently in the two qubit branches and are the
usual driver of echo decay. The crucial point for the method is that we retain the \emph{full} tensor
$\mathsf{A}_i$, including the components $A^{xx}$, $A^{yy}$, $A^{xy}$ that flip
the electron spin. Those are not protected by the mirror: under $z \to -z$ both
$\mathbf{S}$ and $\mathbf{I}$ transform as pseudovectors, so $S_z I_x$ is odd
and vanishes for in-plane nuclei, but $S_x I_x$ and $S_y I_y$ are \emph{even}
and survive. Section~\ref{sec:method} explains why that matters.\\

The three first-shell nuclei carry
measured tensors; everything beyond them is dipolar. For the first shell we use
the ENDOR values of Ref.~\cite{Gracheva2023}, $A_\parallel = 87.0$ and
$A_\perp = 45.5$~MHz for $^{14}$N, in agreement with the first-principles
tensor $(44.97, 46.12, 87.15)$~MHz of Ref.~\cite{Ivady2020}. The essential
feature is that this tensor is axial about the \emph{in-plane} nitrogen
dangling bond and not about the $c$ axis, so the component that enters
$A^{zz}$ is $A_\perp$. Scaling by the gyromagnetic ratio
$\gamma(^{15}\mathrm{N})/\gamma(^{14}\mathrm{N}) = 1.403$ gives
$A^{zz} = 63.8$~MHz for the $^{15}$N of h$^{11}$B$^{15}$N, consistent with the
$64$~MHz measured directly on $^{15}$N-enriched crystals~\cite{Sasaki2023}.

Beyond the first shell we use a point-dipole coupling to a distributed spin
density: weight $0.20$ on each of the three first-shell nitrogens and the
remaining $0.40$ at the vacancy centroid, all in the defect plane. This is what
makes the mirror selection rule exact in our model
(Sec.~\ref{sec:layers}). Quadrupole tensors are axial about $c$ with
$C_q(^{11}\mathrm{B}) = 2.934$~MHz and $\eta \approx 0$ from solid-state
NMR of \hBN{}; in h$^{11}$B$^{15}$N the quadrupole term acts on $^{11}$B alone,
since $^{15}$N has $I = \tfrac12$ and no quadrupole moment. The internuclear
couplings $\mathsf{J}_{ij}$ are point-dipole, with
$\gamma(^{11}\mathrm{B}) = 13.663$ and
$\gamma(^{15}\mathrm{N}) = 4.316$~MHz/T.

\subsection{The coherence function}
\label{sec:coherence}

The expansion of Sec.~\ref{sec:ecs} factorises a single scalar, Eq.~\eqref{eq:ecs}; we define it here.
Prepare $\rho(0) = \lvert \psi \rangle \langle \psi \rvert \otimes \mathcal{I}/d$
with $\lvert \psi \rangle = (\lvert 0 \rangle + \lvert -1 \rangle)/\sqrt{2}$ and
the bath fully mixed, nuclear Zeeman energies being far below $k_B T$. Propagate
under the \emph{full} Eq.~\eqref{eq:H}, electron and nuclei together, with no
conditioning on the qubit state, interrupted by $N$ ideal $\pi$ pulses
$\mathcal{P}$ acting on the electron alone; for $\tau = t/2N$:
\begin{equation}
U_N(t) = U_\tau \left[ \mathcal{P}\, U_{2\tau} \right]^{N-1} \mathcal{P}\, U_\tau ,
\quad U_s = e^{-2\pi i H s}.
\label{eq:useq}
\end{equation}
The coherence is the surviving off-diagonal element of the qubit, traced over
the bath,
\begin{equation}
L(t) \propto \mathrm{Tr}\!\left[ U_N(t)\, \rho(0)\, U_N^{\dagger}(t)\,
\bigl( \lvert -1 \rangle \langle 0 \rvert \otimes \mathcal{I} \bigr) \right] ,
\label{eq:L}
\end{equation}
normalised to $L(0) = 1$. Then $T_2$ is the first time $\lvert L \rvert = 1/e$,
and $\lvert L(t) \rvert$ is what a Hahn echo ($N = 1$) measures. Each factor
$L_X$ in Eq.~\eqref{eq:ecs} is this same quantity computed \emph{exactly}, with
Eq.~\eqref{eq:H} restricted to the central block plus the nuclei in $X$.

\section{Method}
\label{sec:method}

\subsection{Beyond standard cluster expansion}

The cluster-correlation expansion (CCE), introduced by Yang and
Liu~\cite{Yang2008} and Witzel and Das Sarma~\cite{Witzel2006}, is now the
standard tool for nuclear-spin-induced decoherence of solid-state
qubits~\cite{Onizhuk2025}. It computes
decoherence by chopping the bath into small groups, evaluating each group's
effect on the qubit separately, and multiplying the results. In its standard
form the electron is not a dynamical object: it is frozen along $z$ and
merely supplies a static, state-dependent field to the nuclei, so one evolves
the bath twice, once for $m_s = 0$ and once for $m_s = -1$, and the coherence
is the overlap of the two. That is exact if the electron--nuclear coupling is
limited to $S_zI_z$ and $S_zI_\perp$.
This approximation is not adequate for $\VB$. Reference~\cite{Tarkanyi2026} benchmarks
precisely this approximation against the generalized expansion and reports that ``the
pseudo-secular approximation of the hyperfine interaction yields essentially the
CCE method, which, in contrast to gCCE, ignores electron spin flipping
processes,'' and that below $350$~mT the generalized method must be used. We
reproduce that finding independently: with the full tensor, our CCE and gCCE
coherence functions differ by up to $0.79$ at $50$~mT, and the discrepancy
closes to $0.19$ only by $1$~T. In the generalized expansion (gCCE) the electron sits \emph{inside} every
cluster and is propagated with the nuclei, so the full hyperfine tensor acts and
electron--nuclear flip-flops are captured. The cost is a factor $2S+1 = 3$ in
cluster dimension. The calculations reported here use our own implementation,
written from the published formulation of the method and taking
PyCCE~\cite{Onizhuk2021} as a methodological reference. Because the two share
no code, the bulk benchmark of Sec.~\ref{sec:valid} constitutes an independent
comparison.

\subsection{An extended central spin}
\label{sec:ecs}

The method mentioned above is still not sufficient, for a reason specific to this defect. The three
first-shell nitrogens are coupled to the electron at $64$--$122$~MHz. Together
with the electron, they behave as a single strongly entangled object, and any
expansion that puts two of them in one cluster and the third in another discards
correlations that are not small. We see this directly: against exact
diagonalisation of a five-spin bath, plain gCCE gives errors of $0.53$, $0.50$
and $1.26$ at first, second and third order, it gets \emph{worse}, because the
product of cluster corrections diverges when a strongly coupled trio is split.\\

We therefore adopt the structure Cholsuk, Vogl and Iv\'ady found necessary for
the relaxation problem~\cite{IvadyT1}: an indivisible \emph{central block} composed
of the electron and all three first-shell nitrogens, solved exactly, with the
expansion carried out only over the genuinely weakly coupled remainder, that is:
\begin{equation}
\begin{aligned}
L &= L_C \prod_i \tilde L_i \prod_{i<j} \tilde L_{ij}, \\[2pt]
\tilde L_i &= \frac{L_{C+i}}{L_C},
\qquad
\tilde L_{ij} = \frac{L_{C+ij}}{L_C \, \tilde L_i \tilde L_j} .
\end{aligned}
\label{eq:ecs}
\end{equation}
In h$^{11}$B$^{15}$N the central block is only $3 \times 2^3 = 24$ dimensional,
because $^{15}$N is spin-$\tfrac12$, and a cluster with two extra bath spins is
at most $384$ dimensional.

\begin{table*}[t]
\caption{Validation of the extended-central-spin gCCE implementation. The
reduction test shows the new code collapsing onto the old where it must; the
bulk-limit row is an external check against published work.}
\label{tab:valid}
\begin{ruledtabular}
\begin{tabular}{lll}
check & quantity & result \\
\colrule
normalisation & $|L(0)| - 1$ & $0$ (exact) \\
hermiticity & $\max|H - H^\dagger|$ & $0$ (exact) \\
two-nucleus bath & gCCE-2 vs exact diagonalisation & $1\times10^{-16}$ \\
reduction & non-secular $\to 0$: gCCE vs CCE & $3\times10^{-13}$ \\
convergence & extended block, order $0/1/2$ & $0.26 / 0.045 / 10^{-16}$ \\
(for contrast) & plain gCCE, order $1/2/3$ & $0.53 / 0.50 / 1.26$ \\
bath radius & $R = 8$ vs $9.5$~\AA, $10$~mT & $0.2\%$ \\
bath radius & $R = 8$ vs $9.5$~\AA, $300$~mT & $8\%$ (ratio $1.7\%$) \\
cluster order & first vs second, $10$~mT & $0.28\%$ \\
cluster order & first vs second, CPMG-$N$ & $<0.3\%$ \\
time grid & $T_2$ on halving $\delta t$ & $0.00\%$ \\
\textbf{bulk limit} & 5 layers vs Ref.~\cite{Tarkanyi2026} & $\mathbf{199}$ vs $\mathbf{220}\ns$ \\
\end{tabular}
\end{ruledtabular}
\end{table*}

\subsection{Validation}
\label{sec:valid}

Table~\ref{tab:valid} lists the checks that we have performed across the simulations. Three deserve comments. The reduction
test is the strongest internal one: when the non-secular hyperfine components
are set to zero by hand, the generalized code must collapse onto the standard
one, and it does to $10^{-12}$. The convergence test shows the extended
central block reaching machine precision where plain gCCE diverges. And the bulk
benchmark is external: our five-layer slab gives $199\ns$ for bulk
h$^{11}$B$^{15}$N, to be compared with the low-field plateau of $220\ns$
reported in Ref.~\cite{Tarkanyi2026} (quoted there as $\approx 200\ns$), a
$10\%$ agreement between independent implementations with independently
assembled couplings.

\subsection{Convergence and the time grid}
\label{sec:conv}

All results use a bath of radius $R = 8$~\AA. To check this, we recomputed the
two working points at $R = 9.5$~\AA, which raises the bath from $72$ to $108$
spins for a monolayer and from $218$ to $326$ for three layers. At $10$~mT the
coherence moves by $-0.0\%$ and $-0.2\%$ respectively, and the layer ratio by
$+0.2\%$, therefore, converging. At $300$~mT the three-layers envelope is unchanged and the
monolayer envelope falls by $8.0\%$, from $86.5$ to $79.6\us$. (The same
$79.6\us$ appears in Sec.~\ref{sec:res-field} as the monolayer envelope at
$220$~mT; the two are distinct calculations that happen to land on the same
point of the common logarithmic time grid.) That shift is
two steps of the logarithmic time grid, whose spacing is $4.3\%$ per step, so it
is at the resolution of the estimator rather than above it; the threshold-free
mean amplitude, which does not read a grid crossing at all, moves by $-4\%$ and
$-6\%$ and its layer ratio by $+1.7\%$. Our results are assigned with a systematic $8\%$
to absolute high-field coherence times, and none to the layer ratios.\\

Cluster order is the second convergence parameter, and at the working point it
is benign. Recomputing the $10$~mT Hahn echo at both orders with an otherwise
identical pipeline gives $651.5 \to 653.3\ns$ for one layer,
$230.1 \to 230.4\ns$ for two and $198.3 \to 198.8\ns$ for three, that is
$+0.28\%$, $+0.13\%$ and $+0.25\%$. Every figure and table in this paper is
second order at $R = 8$~\AA{} with two exceptions, each noted where it occurs:
the resolved traces of Fig.~\ref{fig:cpmg}(a) and the sublattice comparison of
Fig.~\ref{fig:mechanism}(a), for which both members were computed at first
order. Elsewhere first-order values appear only where an order comparison is the
point. This agreement does \emph{not} extend to the
bottom of the field range, where the two orders diverge from one another
(Sec.~\ref{sec:limits}).

Second order is sufficient at the working point, for a physical reason.
Third-order clusters describe correlated flip-flops among three bath nuclei,
which set the high-field plateau of bulk h$^{11}$B$^{15}$N and are therefore
retained in Ref.~\cite{Tarkanyi2026}. At low field the coherence is instead
governed by the out-of-plane hyperfine of individual borons, i.e., by a
first-order quantity, and the $0.28\%$ separation between first and second order
quantifies this: the correction still being resolved by the expansion is three
orders of magnitude smaller than the effect we report. We therefore expect
third-order terms to be smaller still, and we take the first-to-second-order
interval as our estimate of the convergence error.

The time grid needs more care than usual. Because the $\VB$ echo carries a
deep modulation at the first-shell hyperfine frequency ($\approx 67$~MHz,
period $15\ns$), the $1/e$ crossing can fall inside a narrow modulation minimum,
and a grid that steps over that minimum reports the \emph{next} one instead. We
therefore locate $T_2$ in two passes, i.e., a coarse logarithmic grid to bracket it,
then a uniform grid with $\delta t \le (8 f_{\max})^{-1} \approx 1.8\ns$. We also recompute a sample of points at half the step and record the discrepancy rather than ignoring it. Those checks return $0.00\%$ to the digits carried.

\begin{figure*}[t]
\includegraphics[width=\textwidth]{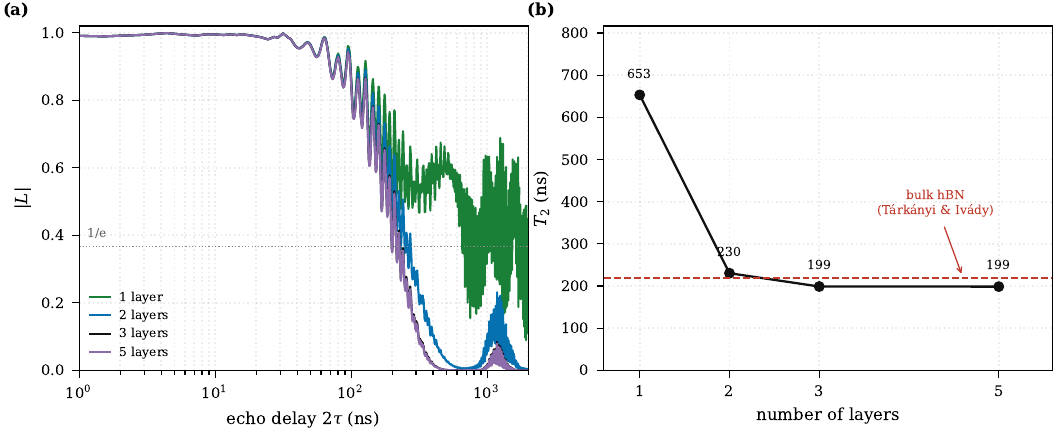}
\caption{Coherence versus layer number in h$^{11}$B$^{15}$N at $B_0 = 10$~mT.
(a) The Hahn echo. The modulation at the first-shell hyperfine frequency is
visible throughout and is why $T_2$ must be extracted on a resolved time grid
(Sec.~\ref{sec:conv}). (b) The resulting $T_2$: the bulk limit is reached by
three layers, and agrees with the published bulk value~\cite{Tarkanyi2026},
computed independently.}
\label{fig:layers}
\end{figure*}

\section{Results}
\label{sec:results}

\subsection{Layer number and sublattice carrying the noise}
\label{sec:res-layers}

Figure~\ref{fig:layers} shows the echo and the resulting coherence time in
h$^{11}$B$^{15}$N at $B_0 = 10$~mT. The monolayer reaches
$T_2 = 653\ns$; a bilayer falls to $230\ns$; three layers give $199\ns$ and five
give the same, so the bulk limit is reached at the third layer. The uniform
second-order sweep sharpens this: throughout the range where the calculation is
quantitative, $2$ to $110$~mT, the three- and five-layer coherences agree to
$0.32\%$ or better at every field, so saturation at three layers is not a
property of the working point but holds across the whole low-field regime.
Sec.~\ref{sec:layers} provides us with an interpretation for that. Going from one layer
to two adds a full sheet of unprotected nuclei $3.33$~\AA{} away; going from two
to three adds a second sheet on the other side, which is already a weaker
perturbation because the first has done most of the damage; beyond that the new
nuclei are too distant to matter.\\

\begin{figure*}[t]
\includegraphics[width=\textwidth]{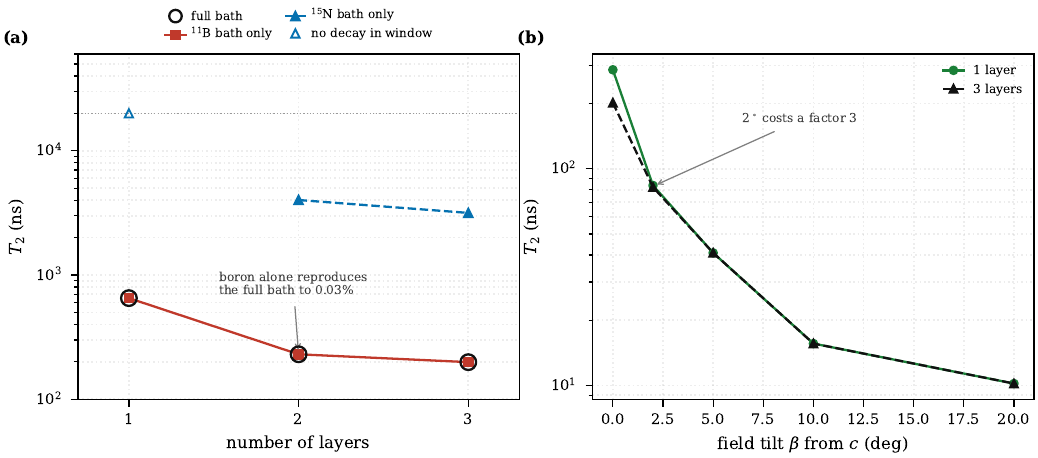}
\caption{(a) Which sublattice carries the noise, at $B_0 = 10$~mT. Removing the
nitrogen bath changes nothing; removing the boron bath changes everything. The
layer-number effect is a boron effect. (b) Fragility against field
misalignment, here at $B_0 = 50$~mT. Tilting $B_0$ by $\beta$ from the $c$ axis
switches on the transverse hyperfine coupling of the three first-shell
nitrogens, forbidden at $\beta = 0$; since those nitrogens are present in every
configuration, the loss is independent of layer number. The tolerance is itself
field dependent: comparison with Fig.~\ref{fig:tilt}, the same sweep at the
$10$~mT working point, gives the $1/B_0$ scaling of Sec.~\ref{sec:res-tilt}.}
\label{fig:mechanism}
\end{figure*}

If the mechanism identified in the bulk drives the layer effect, it must be carried by
boron. Figure~\ref{fig:mechanism}(a) tests this by rebuilding the bath with only
one sublattice present (the three first-shell nitrogens are retained in all
cases, since they are part of the central block). The boron bath alone gives
$651.3, 230.0, 198.9\ns$ for one, two and three layers, against $651.5, 230.1,
198.8\ns$ for the full bath, an agreement of $0.03\%$. (This comparison is at
first order, where both members were computed; the second-order full-bath values
are $653.3, 230.4, 198.8\ns$, and the conclusion is unchanged.) The nitrogen
bath alone
gives no decay at all within our window for the monolayer and $4.0$ and
$3.2\us$ for two and three layers, an order of magnitude longer than the full
result. That demonstrates that the coherence of $\VB$ at low field is set by the boron nuclei; 
the monolayer removes two thirds of the
boron that is close enough to matter, and $T_2$ rises accordingly.

\subsection{Field dependence of the enhancement}
\label{sec:res-field}

\begin{figure*}[t]
\includegraphics[width=\textwidth]{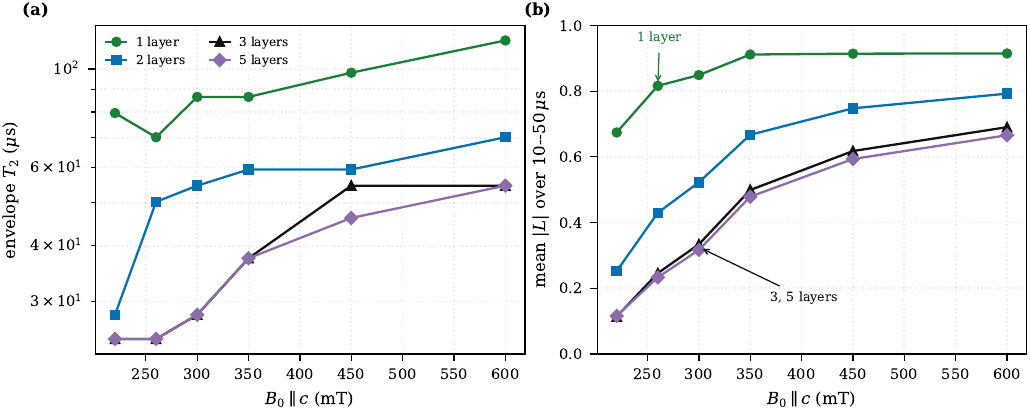}
\caption{High field, second order, $R = 8$~\AA. (a) Envelope coherence, the last
crossing of $1/e$; the first crossing is not usable here
(Sec.~\ref{sec:res-field}). (b) Mean echo amplitude over $10$--$50\us$, which
involves no threshold. The monolayer keeps $0.67$--$0.91$ of its coherence over
this range and three and five layers only $0.11$--$0.69$, the three- and
five-layer curves coinciding to $1$--$5\%$, so the bulk limit is again reached
at three layers. Fields below $200$~mT are absent: there the expansion diverges,
mean $|L|$ exceeds unity near the $124$~mT anticrossing, and the points are
rejected automatically.}
\label{fig:highfield}
\end{figure*}

Figure~\ref{fig:layersfield} follows the three configurations from $1$ to
$1000$~mT, all at second order. The factor of $3.3$ holds from $2$ to about
$25$~mT ($3.37$, $3.29$, $3.29$, $3.29$ and $3.28$ at $2$, $5$, $8$, $10$ and
$25$~mT, respectively), and then closes: by
$85$~mT the three give $162.6$, $145.1$ and $144.2\ns$, a residual contrast of
$1.13$ against the $3.29$ at the working point. At $124$~mT
the ground-state level anticrossing~\cite{Mathur2022,Gao2022NatMater,Ru2024},
where $\lvert 0 \rangle$ and $\lvert -1 \rangle$ become degenerate, collapses
all three identically to $3.4\ns$, as it must, since that is an electronic
degeneracy and has nothing to do with the bath; because it is independent of
the bath and of the expansion, this collapse serves as an internal consistency
check even though the cluster expansion is otherwise unreliable in this field
range (Sec.~\ref{sec:limits}). Above the anticrossing the
enhancement reopens as Fig.~\ref{fig:highfield} shows. At this point two aspects need to be settled first:\\
\emph{The threshold must be shifted.} Beyond $\approx 200$~mT neither
configuration has fully decohered within our window and the echo is strongly
modulated, so the \emph{first} crossing of $1/e$ reports a modulation minimum
rather than a decay: at $350$~mT the coherent oscillation, not the underlying
decay, first carries the echo below the $1/e$ threshold at $0.16\us$ for three
layers, while the slowly decaying envelope retains a mean amplitude of $0.50$
over the $10$--$50\us$ window. We therefore use the \emph{last} crossing of
$1/e$, the envelope, and quote the mean amplitude, which involves no
threshold at all, alongside it.\\
\emph{About mean amplitude field range.} Above the anticrossing at $124$~mT, $T_2$ is insufficiently described; we use a second-order mean $|L|$ and the envelope coherence to witness the layer-number effect. These quantities can diverge when the anticrossing collapses the qubit splitting, which makes the product of the cluster corrections blow up and returns unphysical mean amplitudes; that is the case for one, three and five layers in the vicinity of the anticrossing. We therefore reject any point with mean $|L| > 1$ automatically rather than by inspection, which leaves $220$--$600$~mT as our field range.\\

In that range the result is clean and the bulk limit is again reached at three
layers, the three- and five-layer amplitudes agreeing to $1$--$5\%$. The
monolayer envelope coherence is $79.6\us$ against $24.7\us$ at $220$~mT and
$115.9\us$ against $54.6\us$ at $600$~mT, a factor of $3.2$ falling to $2.1$.
The threshold-free measure tells the same story more smoothly: after
$10$--$50\us$ the monolayer retains $0.67$ of its coherence at $220$~mT rising
to $0.91$ by $350$~mT, while three layers retain $0.11$ rising to $0.50$. The
gap closes monotonically with field, from $0.56$ at $220$~mT to $0.22$ at
$600$~mT.

The lower edge of that range needs one caveat. Cluster order is well converged
down to $2$~mT, where first and second order differ by $2.8\%$ ($650$ against
$668\ns$ for the monolayer), and to $0.3\%$ at $10$~mT and above
(Sec.~\ref{sec:conv}). At $1$~mT it is not: the two orders give $149$ and
$274\ns$, an $84\%$ discrepancy. The breakdown occurs where it should, since there
the electron Zeeman energy $\gamma_e B_0 = 28$~MHz has fallen below the
first-shell hyperfine, the electron is no longer well quantised along $c$, and
the pseudo-secular terms the expansion treats perturbatively cease to be small.
We therefore quote the plateau from $2$~mT upward and exclude the single point
at $1$~mT (Sec.~\ref{sec:limits}); the transverse zero-field splitting excludes
that region on physical grounds as well.

\subsection{Alignment tolerance}
\label{sec:res-tilt}

The protection rests on a mirror symmetry that the applied field must respect,
so it should be fragile against misalignment. It is, but not nearly as fragile
as a sweep taken at the wrong field suggests, and the difference matters for
whether the experiment is easy or hard.

\begin{figure*}[t]
\includegraphics[width=\textwidth]{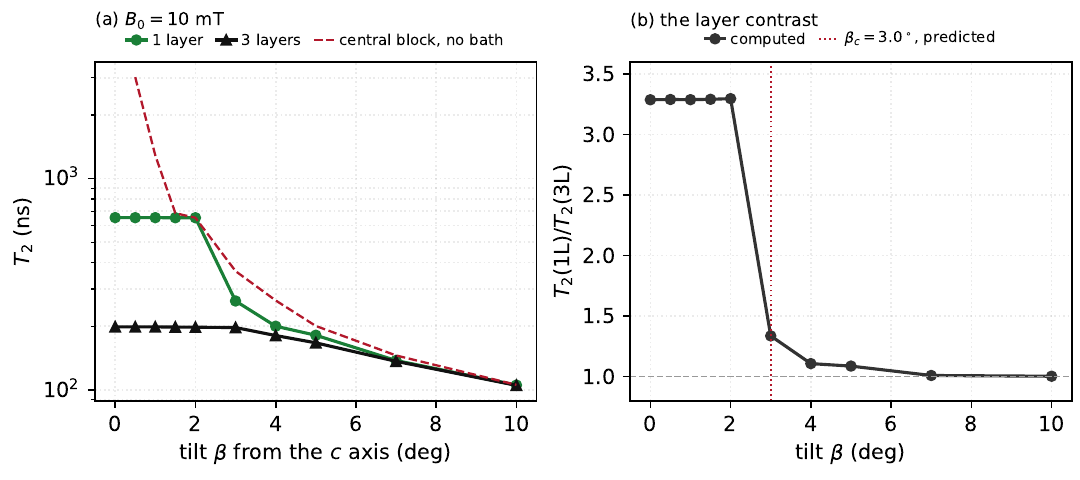}
\caption{Alignment tolerance at the working point, $10$~mT, second order,
$R = 8$~\AA. (a) $T_2$ against tilt. The dashed curve is the same sweep with the
bath removed, leaving only the electron and the three first-shell nitrogens; it
meets the full-bath monolayer at $2^\circ$ and follows it thereafter, so beyond
the onset the coherence is set entirely by the central block. (b) The layer
contrast, flat at $3.29$ to $2^\circ$ and gone by $7^\circ$. Dotted line:
$\beta_c = 3.0^\circ$, predicted with no free parameters from
Eq.~\eqref{eq:azperp}.}
\label{fig:tilt}
\end{figure*}

Figure~\ref{fig:tilt} sweeps the tilt $\beta$ at the working point,
$B_0 = 10$~mT, second order, $R = 8$~\AA. Nothing happens up to $2^\circ$: the
monolayer goes from $653.3$ to $652.4\ns$, the trilayer from $198.8$ to
$198.0\ns$, and the layer contrast holds at $3.29$. Between $2$ and $3^\circ$ it
falls abruptly, to $263.4$ against $197.2\ns$, i.e., a contrast of $1.34$, and by
$7^\circ$ the two thicknesses are indistinguishable. The loss factors relative
to $\beta = 0$ are $1.0$ at $2^\circ$, $2.5$ at $3^\circ$ and $6.2$ at
$10^\circ$.

The reason the tolerance is this generous, and the reason a $50$~mT sweep
misleads, is that the nuclei follow the \emph{electron}, not the applied field.
With $D \gg \gamma_e B_0$ the electron eigenaxis tilts by only
$\theta_e \approx \gamma_e B_0 \sin\beta / D$, so the transverse hyperfine
switched on at the first shell is
\begin{equation}
A^{z\perp} \approx A^{zz}\,\frac{\gamma_e B_0 \sin\beta}{D},
\label{eq:azperp}
\end{equation}
smaller than the naive $A^{zz} \sin\beta$ by the factor $\gamma_e B_0/D$, which
is $0.081$ at $10$~mT. Misalignment therefore matters once
Eq.~\eqref{eq:azperp} reaches the scale of the bath couplings,
$\approx 0.3$~MHz, which happens at $\beta_c = 3.0^\circ$ at $10$~mT and at
$0.6^\circ$ at $50$~mT. Both are what we compute: the cliff in
Fig.~\ref{fig:tilt} sits at $3^\circ$, and the earlier $50$~mT sweep had already
lost a factor of $3.4$ by $2^\circ$. The tolerance scales as $1/B_0$, so
operating at the low-field working point is doubly advantageous, since it is where
the effect is largest \emph{and} where alignment is least demanding.

That the loss is a first-shell effect rather than a bath effect is now
demonstrated rather than inferred. The dashed curve in Fig.~\ref{fig:tilt}(a)
repeats the sweep with the bath deleted entirely, keeping only the electron and
its three first-shell nitrogens. From $\beta = 1.5^\circ$ it tracks the
full-bath monolayer, and at $2^\circ$ the two agree exactly ($652.4\ns$): past
the onset the monolayer coherence is set by the central block alone and the
bath contributes nothing. Since that block is identical in every configuration,
the loss cannot depend on layer number, which is why the one- and three-layer
curves merge.

The practical specification is therefore alignment to within about two degrees,
rather than the fraction of a degree implied by the $50$~mT numbers.
The orientation dependence of $\VB$ coherence has been mapped independently by
Mistri \emph{et al.}~\cite{Mistri2025}, who identify the field configurations
that minimise the energy gradients of the coupled electron--nuclear system.
Their analysis is performed at fixed thickness and ours at fixed orientation, so
that the two are complementary; Eq.~\eqref{eq:azperp} describes the same
competition from the point of view of the defect, and both approaches agree on
the sensitivity of the low-field regime to the field orientation.

\subsection{Registry and twist effects}
\label{sec:res-stack}

Everything so far assumes the AA$'$ registry of bulk \hBN{}. Exfoliated and
transferred flakes carry no such guarantee: parallel-stacked few-layer \hBN{}
sits in the polar AB or BA registry, the two ferroelectric domain states
identified in bilayer \hBN{} by Yasuda et al.~\cite{Yasuda2021} and Vizner Stern
et al.~\cite{ViznerStern2021}, and a transferred flake has an uncontrolled
twist. Twisted bilayers exhibit charge-polarized interfacial superlattices, as
shown by Woods et al.~\cite{Woods2021}. If the prediction of this paper
depended on the stacking, it would be difficult to test. We find that it does not.
Analogous domain networks have been observed in twisted transition metal
dichalcogenide bilayers~\cite{Enaldiev2020}, and recent work has explored
super-moiré domain tessellations in twisted trilayer
\hBN{}~\cite{SuperMoire2026}. A nontrivial constraint predicted by the present work 
is that coherence, as considered within our framework, is insensitive to these arrangements.\\

Table~\ref{tab:stack} gives $T_2$ at $10$~mT for the four high-symmetry
registries. The local neighbourhood changes substantially between them: in AA a
$^{11}$B sits $3.33$~\AA{} directly over the vacancy and the largest
out-of-plane coupling rises from $0.641$ to $0.899$~MHz, a change of $40\%$. The
coherence moves by $0.09\%$ in a bilayer and $0.16\%$ in a trilayer. Rotating
the flanking layers, by $\pm\theta$ for layers $\pm 1$, changes $T_2$ by
$0.16\%$ over the whole range $\theta = 0$ to $30^\circ$, including the
commensurate $21.787^\circ$. The two ferroelectric states are indistinguishable:
AB and BA differ by $0.09\%$ in a bilayer, and in a uniformly slid trilayer they
are exactly degenerate, since layers $\pm 1$ are displaced by $\pm\bm{\tau}$ and
the two configurations are related by $z \to -z$, a symmetry of the defect.\\

\begin{table}[t]
\caption{$T_2$ in ns at $10$~mT against interlayer registry, second order,
$R = 8$~\AA. ``Overhead'' is the nucleus on the $c$ axis through the vacancy
in the layer above; in AB a hexagon centre lies there, the nearest nucleus
being $1.45$~\AA{} off axis. $A^{zz}_{\max}$ (MHz), the largest out-of-plane
hyperfine component, is quoted for the bilayer, where the registry spread is
widest; in the trilayer it is $0.899$ for AA and $0.694$ for the other three. A
$40\%$ change in the strongest single coupling moves $T_2$ by $0.09\%$.}
\label{tab:stack}
\begin{ruledtabular}
\begin{tabular}{ccccc}
registry & overhead & $A^{zz}_{\max}$ & 2 layers & 3 layers \\
\colrule
AA$'$ & $^{15}$N & $0.694$ & $230.4$ & $198.8$ \\
AA & $^{11}$B & $0.899$ & $230.2$ & $198.6$ \\
AB & --- & $0.694$ & $230.4$ & $198.5$ \\
BA & $^{15}$N & $0.641$ & $230.2$ & $198.5$ \\
\end{tabular}
\end{ruledtabular}
\end{table}

\begin{table}[t]
\caption{$T_2$ (ns) under CPMG-$N$ at $B_0 = 10$~mT, second order, $R = 8$~\AA.
These are first $1/e$ crossings, which under CPMG read a modulation minimum
rather than a decay (Fig.~\ref{fig:cpmg}); the threshold-free contrast is in
Fig.~\ref{fig:cpmg}(b). First order reproduces every entry to better than
$0.3\%$. $N = 8$ is omitted: the expansion diverges there
(Sec.~\ref{sec:limits}).}
\label{tab:dd}
\begin{ruledtabular}
\begin{tabular}{ccccc}
$N$ & 1 layer & 2 layers & 3 layers & 1L/3L \\
\colrule
1 & $653.3$ & $230.4$ & $198.8$ & $3.29$ \\
2 & $206.3$ & $206.3$ & $206.3$ & $1.00$ \\
4 & $283.2$ & $223.6$ & $223.6$ & $1.27$ \\
\end{tabular}
\end{ruledtabular}
\end{table}

A rotation about the vacancy preserves the distance \emph{and} the polar angle of
every nucleus; only the azimuth changes. For a point spin density at the vacancy,
$A^{zz}$ would be exactly invariant, and the residual $0.16\%$ is the small
correction from the weight carried on the three first-shell nitrogens, which are
themselves $C_3$-symmetric about the same axis. A slide is not a symmetry
operation and does move nuclei, by up to $1.45$~\AA{} in plane, however, the
flanking layer remains a complete \hBN{} sheet at the same height, with the same
areal density of spins, and the number of out-of-plane nuclei inside the bath is
unchanged at $73$ per flanking layer. The decoherence is a collective effect of
those $\sim 70$ spins.\\

The controlling variable is therefore the number of spin-carrying planes and
their separation from the defect, rather than the in-plane arrangement of those
planes. Even the separation enters weakly: changing it by $0.6$~\AA{} alters the
strongest out-of-plane coupling by $32\%$ and $T_2$ by only $15\%$
(Sec.~\ref{sec:limits}), so the collective average absorbs about half of that
too. This has two implications. Experimentally, the proposed measurement
requires no registry or twist control, which removes what would otherwise be a
demanding experimental constraint (the hardest one that remains is the field
alignment of Sec.~\ref{sec:res-tilt}). On the other hand, the same insensitivity
implies that $\VB$ coherence does not carry a signature of the ferroelectric
polarization, the AB and BA registries differing by only $0.09\%$ in a bilayer
and being exactly degenerate in a uniformly slid trilayer for the reason given
above.

\begin{figure}[t]
\includegraphics[width=\columnwidth]{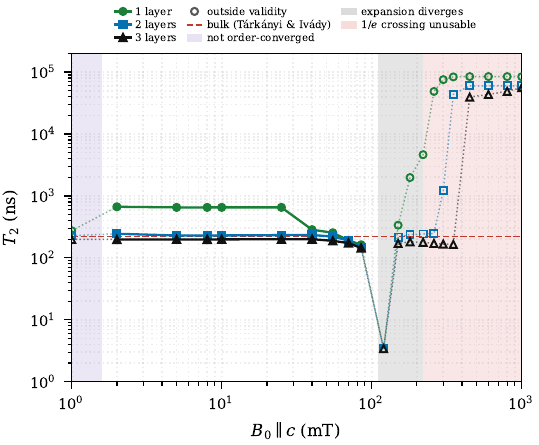}
\caption{$T_2$ versus applied field for one, two and three layers, second
order, $R = 8$~\AA{} throughout. The layer-number enhancement is a low-field
effect: a factor of $3.3$ from $2$ to $25$~mT, largely closed by $85$~mT, and
zero at the ground-state level anticrossing at $D/\gamma_e = 124$~mT, where all
thicknesses collapse to $3.4\ns$. Filled symbols are quantitative; open symbols mark the
three shaded regions where they are not. Lavender, $1$~mT: not converged in cluster order, first and second order
differing by $84\%$. Grey, $110$--$220$~mT: the expansion diverges near the
anticrossing. Pink, above $220$~mT: the echo is modulated but barely decayed, so
the first $1/e$ crossing reads a modulation minimum and understates the
coherence by orders of magnitude ($167\ns$ against an envelope of $28\us$,
three layers at $300$~mT); use the envelope estimator of
Fig.~\ref{fig:highfield}. Red dashed line: the bulk value
of Ref.~\cite{Tarkanyi2026}.}
\label{fig:layersfield}
\end{figure}

\subsection{Dynamical decoupling and the danger of a threshold}
\label{sec:res-dd}

If the layer-number effect originates in bath noise, a pulse sequence that
refocuses the bath should weaken it. Testing this expectation requires some care
in what is meant by a coherence time.

Table~\ref{tab:dd} lists $T_2$ under CPMG-$N$ at $10$~mT from the same two-pass
resolved-grid estimator used everywhere else in this paper, called with $N$
$\pi$ pulses instead of one. The Hahn echo shows the factor of $3.29$ that is
the subject of this work. At $N = 2$ all three thicknesses return $206.3\ns$,
agreeing to four significant figures, which suggests a complete collapse of the
layer contrast.

\begin{figure*}[t]
\includegraphics[width=\textwidth]{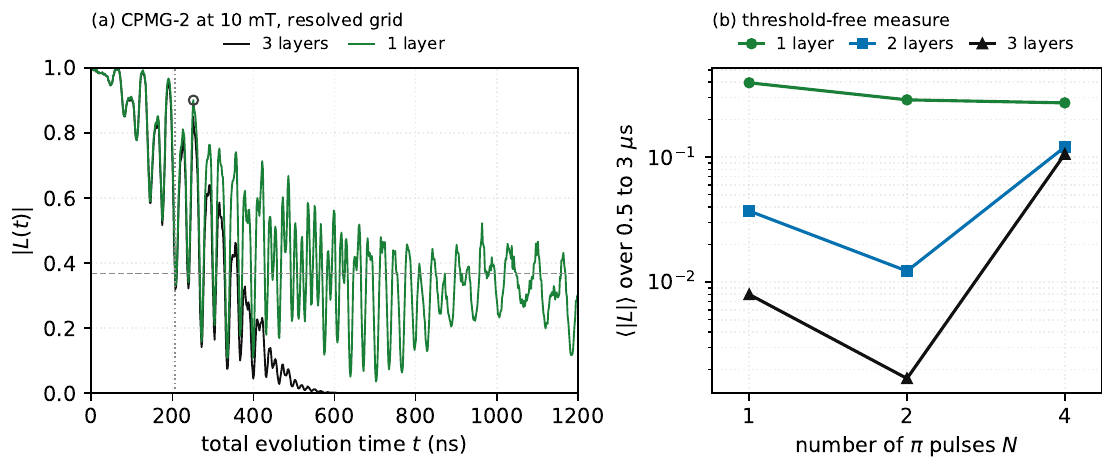}
\caption{Why the CPMG-2 coherence time is not a coherence time. (a) $|L(t)|$
under CPMG-2 at $10$~mT on a uniform $1.7\ns$ grid, first order; this is the one
panel not computed at second order, the two agreeing to $0.3\%$ here
(Sec.~\ref{sec:conv}). Both thicknesses cross $1/e$ at $207\ns$ and both recover
immediately: the crossing is a minimum of the first-shell modulation, common to
every thickness because the central block is. Beyond it the curves diverge, the
trilayer having decayed by $600\ns$ while the monolayer persists past
$1.2\us$. (b) The same physics without a threshold: time-averaged amplitude over $0.5$--$3\us$, second order, which
preserves a contrast between one and three layers of $49$ at $N = 1$, $170$ at
$N = 2$ and $2.6$ at $N = 4$.}
\label{fig:cpmg}
\end{figure*}

This collapse is, however, an artefact of the threshold, as Fig.~\ref{fig:cpmg}(a)
indicates. Resolving $|L(t)|$ under CPMG-2 on a uniform $1.7\ns$ grid, both
thicknesses do cross $1/e$ at $207\ns$, but both recover immediately, the
monolayer to $0.90$ and the trilayer to $0.85$ by $252\ns$. The crossing is a
$45\ns$-wide minimum of the coherent first-shell modulation, not a decay. It is
common to every thickness for the same reason the modulation is: the three
first-shell nitrogens sit inside the central block and are identical by
construction in a monolayer and in the bulk. A crossing set by bath noise could
not be shared by baths of $73$ and $219$ spins.

Beyond that minimum the two curves separate completely. The trilayer is gone by
$600\ns$; the monolayer is still oscillating about $0.3$ at $1.2\us$. Measured
without a threshold, by the time-averaged amplitude $\langle |L| \rangle$ over
$0.5$--$3\us$, the contrast does not collapse at all (Fig.~\ref{fig:cpmg}(b)):
one against three layers gives $0.394$ against $0.008$ at $N = 1$, $0.288$
against $0.0017$ at $N = 2$, and $0.272$ against $0.106$ at $N = 4$. Far from
erasing the layer dependence, two decoupling pulses give the largest separation
we compute. The resolved first-order curves of Fig.~\ref{fig:cpmg}(a) reproduce
these averages to within the order uncertainty ($0.382$ and $0.0079$ at
$N = 1$; $0.327$ and $0.0017$ at $N = 2$), so they are not an artefact of the
sampling either.\\

We note that this observation is not specific to the present system. A $1/e$
crossing is a coherence time only when the signal decays monotonically. For a defect whose echo carries deep
modulation from strongly coupled nuclei inside the qubit, $\VB$ being an
extreme case with three $64$~MHz nitrogens, the crossing can be pinned by a
modulation zero that is a property of the central spin system and carries no
information about the environment at all. We were forced to the same conclusion
independently at high field (Sec.~\ref{sec:res-field}). Reported coherence
times for $\VB$ under decoupling should state which feature of the echo the
threshold is reading.\\

For the experiment, the layer contrast is \emph{not} specific
to the Hahn echo, so a decoupling sequence does not destroy the signal one
would look for; but it must be read from the echo envelope or a time-averaged
amplitude rather than from a first $1/e$ crossing, which under CPMG returns
$206\ns$ for every thickness. And our range is limited: the expansion diverges
beyond $N = 4$ (Sec.~\ref{sec:limits}), so we cannot address the large-$N$
regime in which pulsed decoupling extends $\VB$ coherence in natural \hBN{}
from $58.5\ns$ to $4.2\us$ at $N = 1000$~\cite{Rizzato2023}, or in which
continuous driving achieves a comparable extension~\cite{Ramsay2023}. Nothing
here contradicts those measurements; what Table~\ref{tab:dd} shows is that a
threshold-based $T_2$ under decoupling can be blind to a factor of $170$ in
retained coherence.

\section{Limits of validity}
\label{sec:limits}

Several boundaries delimit the scope of our calculation, each of which we state explicitly.

\emph{The cluster expansion fails near the ground-state level anticrossing.}
At $150$ and $180$~mT the computed mean amplitude exceeds unity, by factors of up
to $2\times10^4$, and grows with the number of layers, that is, with the
number of clusters in the product. This is the standard failure of a cluster
expansion at a level anticrossing: at $124$~mT the qubit splitting collapses to a
few MHz, comparable to the first-shell hyperfine, and the corrections cease to be
small. We exclude these fields rather than interpret them, and note that
$T_2$ from Fig.~\ref{fig:layersfield} between $110$ and $200$~mT inherits the
same problem. Collecting the boundaries of this section, the calculation is
quantitative for $2\text{--}110$~mT and, with the envelope estimator, for
$220\text{--}600$~mT; it is not to be trusted at $1$~mT or between $110$ and
$220$~mT, and above $220$~mT the first $1/e$ crossing must not be used. A treatment valid across the anticrossing would require
diagonalising the electron--nuclear manifold without a cluster expansion. Near
the GSLAC, Fig.~\ref{fig:layersfield} shows that all three thicknesses converge
to the same $T_2$, as expected since the electronic degeneracy at the
anticrossing is independent of the bath. This convergence serves as a useful
consistency check for the calculation.\\

\emph{The stacks are idealised.} Section~\ref{sec:res-stack} varies the registry
and the twist angle, but always with unrelaxed layers. Real twisted and slid
stacks relax, and in the ferroelectric case the relaxation is itself the physics
of interest~\cite{Cortes2023}. Since we find the coherence insensitive to
in-plane rearrangements of a full sheet at fixed height, we expect relaxation to
matter only through the interlayer separation, to which the dipolar coupling is
cubically sensitive. We have not computed that dependence, but it is a natural
extension of this work.\\

\emph{Dynamical decoupling is limited to $N \leq 4$ pulses.} The divergence
described above is not confined to the anticrossing. Under CPMG the coherence is
a product of ever more cluster corrections, and at $10$~mT the computed $|L|$
first exceeds unity inside the fitting window at $N = 8$, by $3.6\%$, $4.8\%$
and $11\%$ for one, two and three layers, growing with both $N$ and the number of
clusters, and reaching macroscopic values by $N = 16$. Table~\ref{tab:dd}
therefore stops at $N = 4$. This is a property of the expansion, not of the
defect. It still allows us to establish that the threshold-based $T_2$
equalises across thicknesses while the envelope contrast survives
(Sec.~\ref{sec:res-dd}), but the large-$N$ regime, in which pulsed decoupling
extends $\VB$ coherence experimentally, remains out of reach.\\

\emph{The expansion is not order-converged at $1$~mT.} At $10$~mT and above,
first and second order agree to $0.3\%$ (Sec.~\ref{sec:conv}); at $2$~mT they
differ by $2.8\%$, which we accept, and at $1$~mT by $84\%$ ($149$ against
$274\ns$ for the monolayer), which we do not. The cause is physical rather than numerical: at $1$~mT the electron
Zeeman energy $\gamma_e B_0 = 28$~MHz falls below the first-shell hyperfine, the
electron ceases to be well quantised along $c$, and the pseudo-secular terms
that the expansion treats as a correction are no longer small. This is the same
breakdown that Ref.~\cite{Tarkanyi2026} documents for the standard expansion,
appearing here one order higher. The three curves of
Fig.~\ref{fig:layersfield} at $1$~mT are therefore excluded rather than
interpreted.

\begin{figure*}[t]
\includegraphics[width=\textwidth]{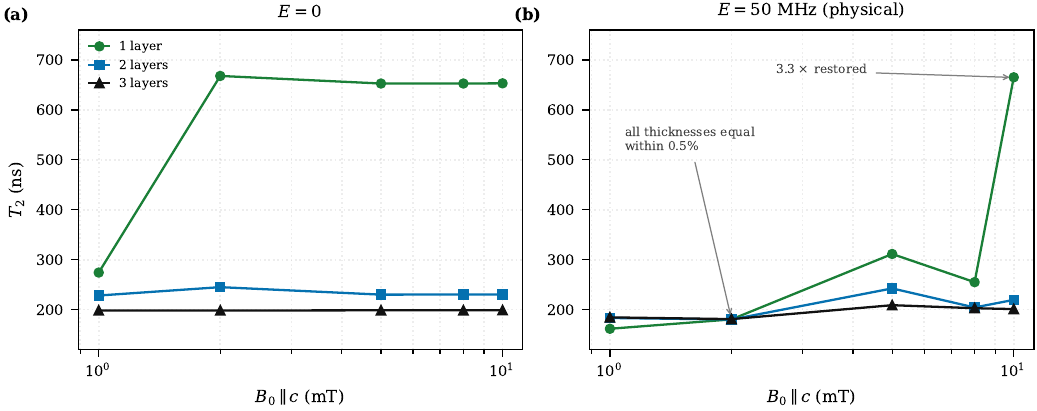}
\caption{Effect of the transverse zero-field splitting at low field, second
order, $R = 8$~\AA. (a) $E = 0$, as Figs.~\ref{fig:layers}--\ref{fig:layersfield}
are computed. (b) $E = 50$~MHz, the physical value. At $2$~mT the three
thicknesses are equal to within $0.5\%$: $\gamma_e B_0 \approx E$ puts the defect at a clock
transition, decoupling it from the bath to first order, so that layer number
stops mattering. By $10$~mT the factor of $3.3$ is fully restored and agrees
with the $E = 0$ value to $2\%$. The intermediate points at $5$ and
$8$~mT lie in the crossover and are not monotonic.}
\label{fig:ezfs}
\end{figure*}

\emph{The enhancement requires $B_0 \gtrsim 10$~mT.} Figure~\ref{fig:ezfs}
compares the low-field sweep with the transverse zero-field splitting set to
$E=0$ and to its physical value $E \approx 50$~MHz~\cite{Mathur2022}. At
$2$~mT all three
thicknesses collapse onto one another, $180.4$, $180.2$ and $181.0$~ns for
one, two and three layers, a spread of $0.5\%$, because $\gamma_e B_0 \approx
E$ there and the defect sits at a clock transition, where it is decoupled from
the bath to first order and thickness ceases to matter. No net protection
accompanies this equalisation: the coherence at the crossing does not exceed
the bulk plateau, consistent with Ref.~\cite{Tarkanyi2026}, which finds that
the clock-transition enhancement expected at $\gamma_e B_0 \approx E$ is
absent for $\VB$ because the strong first-shell hyperfine opens
$\Delta m_s = \pm 2$ channels that bypass the first-order protection. Between $5$ and $8$~mT
the effect is partially restored and non-monotonic (ratios $1.5$ and $1.3$), a
crossover we do not attempt to resolve. By $10$~mT it is fully back: $665.4$~ns
against $201.0$~ns, a factor of $3.31$, compared with $653.3$ and $198.8$~ns and
a factor of $3.29$ at $E = 0$. The working point of this paper is therefore
unaffected by $E$ at the $2\%$ level, though the enhancement should not be sought
below $10$~mT. Figures~\ref{fig:layers}--\ref{fig:layersfield} are computed at
$E = 0$; Fig.~\ref{fig:ezfs} quantifies the correction explicitly. Measurements
on few-layer flakes find, in addition, that $E$ itself is strongly reduced with
thickness~\cite{Durand2023}, so this low-field boundary will shift with layer
number, in the favourable direction of lower fields.\\

\emph{Surfaces and substrates are absent.} Room-temperature measurements on
few-layer flakes already show the spin-lattice relaxation time dropping from
$\approx 13$ to $\approx 1\us$ by three to five layers, a degradation
attributed to surface-related noise that freezes out at low
temperature~\cite{Durand2023}; a monolayer is maximally exposed, and on a
substrate the nuclei and paramagnetic defects of the support take the place of
the boron sheets whose removal produces the enhancement. Our numbers therefore
bound the nuclear contribution: the clean comparison is between flakes with
equivalent surface preparation, suspended or encapsulated in a
nuclear-spin-quiet environment, or at temperatures where the surface channel
is frozen out. We also note that $\VB$ has been measured down to three to five
layers but not yet in a true monolayer, where the stability of the charge
state remains to be established~\cite{Durand2023}.\\

\emph{Only nuclear decoherence is included, from an unpolarised bath.} The
nuclear spins are taken at infinite temperature, which is the relevant limit at
thermal equilibrium; optical pumping can polarise the $\VB$ nuclear
environment~\cite{Gao2022NatMater}, and a polarised bath would suppress the
flip-flops that drive the decay, so that our numbers are conservative in this
respect as well. Spin--lattice relaxation, paramagnetic defects and pulse errors are
also absent, and in ensembles the mutual coupling of nearby defects contributes
a further channel~\cite{Rizzato2023}; every number quoted here is therefore an
upper bound. Acoustic-phonon-driven relaxation, studied recently for the $\VB$
centre~\cite{Adhikary2026}, is not included; our numbers therefore apply in the
regime where spin-lattice relaxation is slow compared with the echo timescale.
This omission does not bias the comparison made here: first-principles
spin--phonon calculations give values of $T_1$ in a monolayer and in
AA$'$-stacked \hBN{} that are nearly identical at room
temperature~\cite{Estaji2025}, so that the layer-number dependence we find in
$T_2$ is not offset by a compensating change in $T_1$.
Excited-state effects, such as Jahn--Teller distortion, are outside the scope of
this work~\cite{Benedek2026}. These other ingredients define clear directions
for future investigations rather than diminishing the conclusions drawn here.

\section{Conclusions}
\label{sec:conclusions}

In this work we have shown that the Hahn-echo coherence of the $\VB$ centre in
\hBN{} is not a bulk intrinsic property but depends on the number of layers
present. In h$^{11}$B$^{15}$N at low field, a monolayer reaches
$T_2 = 653\ns$, a factor of $3.3$ longer than the bulk limit of $199\ns$, and
the enhancement is already saturated by three layers. The mechanism behind this
improvement can be traced to a single source: the coherence-limiting noise at low field is carried
entirely by the boron nuclei in the layers immediately flanking the defect plane.
Removing those layers, as in a monolayer, removes the dominant noise source.
This is consistent with the bulk analysis of T\'ark\'anyi and Iv\'ady, and our
sublattice decomposition demonstrates it directly.\\

The enhancement is confined to low fields: it holds from $2$~mT at $E = 0$, or
from $10$~mT once the transverse zero-field splitting is taken into account,
which is also where the expansion becomes order-converged
(Sec.~\ref{sec:limits}), up to $25$~mT, closes by $85$~mT, and reopens above the
ground-state level
anticrossing with a factor of $2$--$3$ from $220$ to $600$~mT. It requires the field aligned to within about two degrees, an onset set by the
tilt of the electron rather than of the field and therefore loosening as $1/B_0$, and
it survives dynamical decoupling: at two CPMG pulses every thickness returns
the same $206\ns$ from a $1/e$ threshold, but that number is a zero of the
first-shell modulation, and the time-averaged amplitude over the same interval
differs by a factor of $170$ between one and three layers.\\

A less expected result is the insensitivity of $T_2$ to the stacking registry
and to the twist angle: changes of up to $40\%$ in the strongest out-of-plane
hyperfine coupling produce at most $0.16\%$ variation in coherence. This follows
from the collective nature of the decoherence, which involves the $73$ nuclei
per flanking layer, and it has the practical consequence that the predicted
enhancement can be observed without stacking control. Conversely, it also
implies that, within the nuclear channel computed here, $\VB$ coherence
carries no signature of the ferroelectric domain state; electric-field (Stark)
shifts of $D$ and $E$, outside our model, remain a possible spectroscopic
signature.\\

Layer number thus emerges as a design parameter for $\VB$-based sensing, one
that is unique to a van der Waals host. The experimental test is conceptually
simple: measure $T_2$ on exfoliated flakes of controlled thickness in
isotopically purified material (all four isotopic compositions of \hBN{},
h$^{11}$B$^{15}$N included, have been grown as bulk
crystals~\cite{Janzen2024}), at a field of $10$--$25$~mT with alignment
better than two degrees, using a Hahn echo. Our prediction is a monotonic
$3.3\times$ improvement from bulk to monolayer, saturating between one and three
layers. Under CPMG the same contrast is present and in fact larger, but it
must be read from the echo envelope: a first $1/e$ crossing returns $206\ns$
irrespective of thickness and would hide the effect entirely.\\

It should be noted that our calculation includes only nuclear-spin-induced decoherence; other
relaxation channels, such as acoustic-phonon-driven spin--lattice relaxation,
paramagnetic defects, and pulse errors, are absent. These omissions define
important directions for future work. Nevertheless, the boundaries are clear,
and the physics we have identified, the layer-number control of coherence
through the collective boron bath, is robust within them. This work, therefore,
establishes layer number as a design parameter for the coherence of a van der
Waals spin qubit, and we hope it motivates the corresponding experimental
effort.

\begin{acknowledgments}
E.S.M. acknowledges financial support from Proyecto Interno USM 2026 PI\_LIR\_26\_12.
J.M.F. thanks Universidad Central de Chile.
\end{acknowledgments}

\section*{Data availability}
The data that support the findings of this article are available from the
authors upon reasonable request.

\section*{AI-use disclosure}
An AI assistant (Claude, Anthropic) was used to assist with structuring and
language editing of the manuscript and with the development of the simulations framework,
plotting and validation code. Calculations were performed using computational resources 
of Grupo de Simulaciones at UTFSM. All AI-assisted output was
reviewed by the authors. All physical content, derivations, and numerical
results were produced and verified by the authors, who take full
responsibility for the final text.

\bibliography{refs}

\end{document}